\documentclass[prm,twocolumn,superscriptaddress]{revtex4-1}
\usepackage[utf8]{inputenc}
\usepackage{xcolor}
\usepackage{amsmath}
\usepackage{graphicx}
\usepackage{booktabs}

\begin{document}
\title{Observation of van der Waals phonons in the single-layer cuprate (Bi,Pb)$_2$(Sr,La)$_2$CuO$_{6+\delta}$}
\author{Y.~Y.~Peng}
\email{yingying.peng@pku.edu.cn}
\affiliation{International Center for Quantum Materials, School of Physics, Peking University, Beijing 100871, China}
\author{I.~Boukahil}
\affiliation{Stanford Institute for Materials and Energy Sciences, SLAC National Accelerator Laboratory, 2575 Sand Hill Road, Menlo Park, California 94025, USA}
\affiliation{Department of Physics, Stanford University, Stanford, California 94305, USA}
\author{K.~Krongchon}
\affiliation{Department of Physics and Materials Research Laboratory, University of Illinois, Urbana, IL 61801, USA}
\author{Q.~Xiao}
\affiliation{International Center for Quantum Materials, School of Physics, Peking University, Beijing 100871, China}
\author{A.~A.~Husain}
\affiliation{Department of Physics and Materials Research Laboratory, University of Illinois, Urbana, IL 61801, USA} 
\author{Sangjun Lee}
\affiliation{Department of Physics and Materials Research Laboratory, University of Illinois, Urbana, IL 61801, USA}
\author{Q.~Z.~Li}
\affiliation{International Center for Quantum Materials, School of Physics, Peking University, Beijing 100871, China}
\author{A.~Alatas}
\affiliation{Advanced Photon Source, Argonne National Laboratory, Argonne, Illinois 60439, USA}
\author{A.~H.~Said}
\affiliation{Advanced Photon Source, Argonne National Laboratory, Argonne, Illinois 60439, USA}
\author{H.~T.~Yan}
\affiliation{Beijing National Laboratory for Condensed Matter Physics, Institute of Physics, Chinese Academy of Sciences, Beijing 100190, China}
\author{Y.~Ding}
\affiliation{Beijing National Laboratory for Condensed Matter Physics, Institute of Physics, Chinese Academy of Sciences, Beijing 100190, China}
\author{L.~Zhao}
\affiliation{Beijing National Laboratory for Condensed Matter Physics, Institute of Physics, Chinese Academy of Sciences, Beijing 100190, China}
\author{X.~J.~Zhou}
\affiliation{Beijing National Laboratory for Condensed Matter Physics, Institute of Physics, Chinese Academy of Sciences, Beijing 100190, China}
\author{T.~P.~Devereaux}
\affiliation{Stanford Institute for Materials and Energy Sciences, SLAC National Accelerator Laboratory, 2575 Sand Hill Road, Menlo Park, California 94025, USA}
\author{L.~K.~Wagner}
\affiliation{Department of Physics and Materials Research Laboratory, University of Illinois, Urbana, IL 61801, USA} 
\author{C.~D.~Pemmaraju}
\affiliation{Stanford Institute for Materials and Energy Sciences, SLAC National Accelerator Laboratory, 2575 Sand Hill Road, Menlo Park, California 94025, USA}
\author{P.~Abbamonte}
\email{abbamont@illinois.edu}
\affiliation{Department of Physics and Materials Research Laboratory, University of Illinois, Urbana, IL 61801, USA} 

\date{\today}

\begin{abstract}

Interlayer van der Waals (vdW) coupling is generic in two-dimensional materials such as graphene and transition metal dichalcogenides, which can induce very low-energy phonon modes. Using high-resolution inelastic hard x-ray scattering, we uncover the ultra-low energy phonon mode along the Cu-O bond direction in the high-$T_c$ cuprate (Bi,Pb)$_2$(Sr,La)$_2$CuO$_{6+\delta}$ (Bi2201). This mode is independent of temperature, while its intensity decreases with doping in accordance with an increasing c-axis lattice parameter. We compare the experimental results to first-principles density functional theory simulations and identify the observed mode as a van der Waals phonon, which arises from the shear motion of the adjacent Bi-O layers. This shows that Bi-based cuprate has similar vibrational properties as graphene and transition metal dichalcogenides, which can be exploited to engineer novel heterostructures.
\end{abstract}

\maketitle

\section{Introduction}

Two-dimensional (2D) materials, such as graphene, transition metal dichalcogenides (TMDs) and strongly correlated materials like iron-chalcogenides and cuprates \cite{2DmaterialsRev} are held together by van der Waals (vdW) forces between different layers. They have attracted intensive interest due to a wealth of electronic properties and excellent mechanical properties, spanning a full range of electronic properties from insulators, semiconductors, metals to superconductors. More intriguingly, due to the weak interlayer vdW forces one can create many exotic heterostructures by control of individual layers, such as twisted bilayer graphene \cite{twistedGraphene}, which exhibits superconductivity and other strongly correlated states. Therefore, vdW materials not only provide a platform to study numerous exotic physical phenomena, but also show great promise for applications, such as optoelectronics, spin- and valley-tronics \cite{2DmaterialsRev,2DApp}.
  
The Bi-based cuprate family is one of the most studied materials among cuprates. Due to the weak vdW interaction within the adjacent Bi-O layers, it can be easily cleaved to get a clean and smooth surface, being widely studied by many surface-sensitive technologies, such as angle-resolved photoemission spectroscopy (ARPES) and scanning tunneling microscopy (STM). There is a rich set of phases in cuprates such as Mott insulator, superconductor, pseudogap and charge order \cite{KeimerNature}. Among them, charge order (CO) is considered as a generic feature in high-$T_c$ cuprates, while important questions remain about the extent to which the charge order influences lattice and charge degrees of freedom. These questions are intimately connected to the origin of the CO and its relation to superconductivity. Typically, CO due to the modulation of electron density is associated with a lattice distortion if the electron-phonon coupling is strong enough. Indeed, due to the presence of CO some phonon anomalies including energy softening and width broadening have been observed at CO wave vector for acoustic and optical phonons in YBa$_2$Cu$_3$O$_6$ (YBCO), (La,Ba)$_2$CuO$_4$ (LBCO) and Bi$_2$Sr$_2$CaCu$_2$O$_8$ (Bi2212) \cite{TaconYBCO,miao2018incommensurate,HePRB}, indicating the importance of lattice dynamics for electronic properties in cuprates. 

In the present work, by using high-resolution inelastic hard x-ray scattering (IXS) \cite{Sinn_2001,Burkel_2000} we studied the low-energy phonons in single-layer cuprate (Bi,Pb)$_2$(Sr,La)$_2$CuO$_{6+\delta}$ (Bi2201) as a function of momentum, temperature and doping. For the first time we observe the low-energy phonons ($\sim$4 meV) originating from the interlayer van der Waals interactions (vdW phonons) in cuprates. The experimental phonon energies agree well with first principles density functional theory (DFT) calculations \cite{HK1964,KS1965}. Our results demonstrate that Bi-based cuprates have a similar interlayer vibration property as other 2D materials like graphene and TMDs, where low-energy interlayer shear and breathing modes ($\sim$2-5 meV) have been observed \cite{ShearModesGraphene,ShearModes}. These low-energy phonons usually play an important role for thermal and electronic conduction properties in 2D van der Waals materials. For example, the ultra-high room-temperature thermal conductivity in graphite is related to phonon hydrodynamics contributions from the out-of-plane momentum of phonons \cite{machida2020phonon}. On the other hand, we observe that the linewidth broadening of the low-energy phonon only occurs in underdoped Bi2201 where it persists up to 300K. This behavior is different from the previous IXS results of double-layer Bi2212 which found low-energy phonon width broadening exists in both underdoped and overdoped samples \cite{HePRB}. This shows that the low-energy phonon broadening in Bi-based cuprates depends on the material-specific details, suggesting an origin in terms of phonon hybridization rather than charge ordering. 

\section{Methods}

\subsection{IXS experiment} 

\begin{figure}[htbp]
    \includegraphics[width=0.85\columnwidth,angle=0]{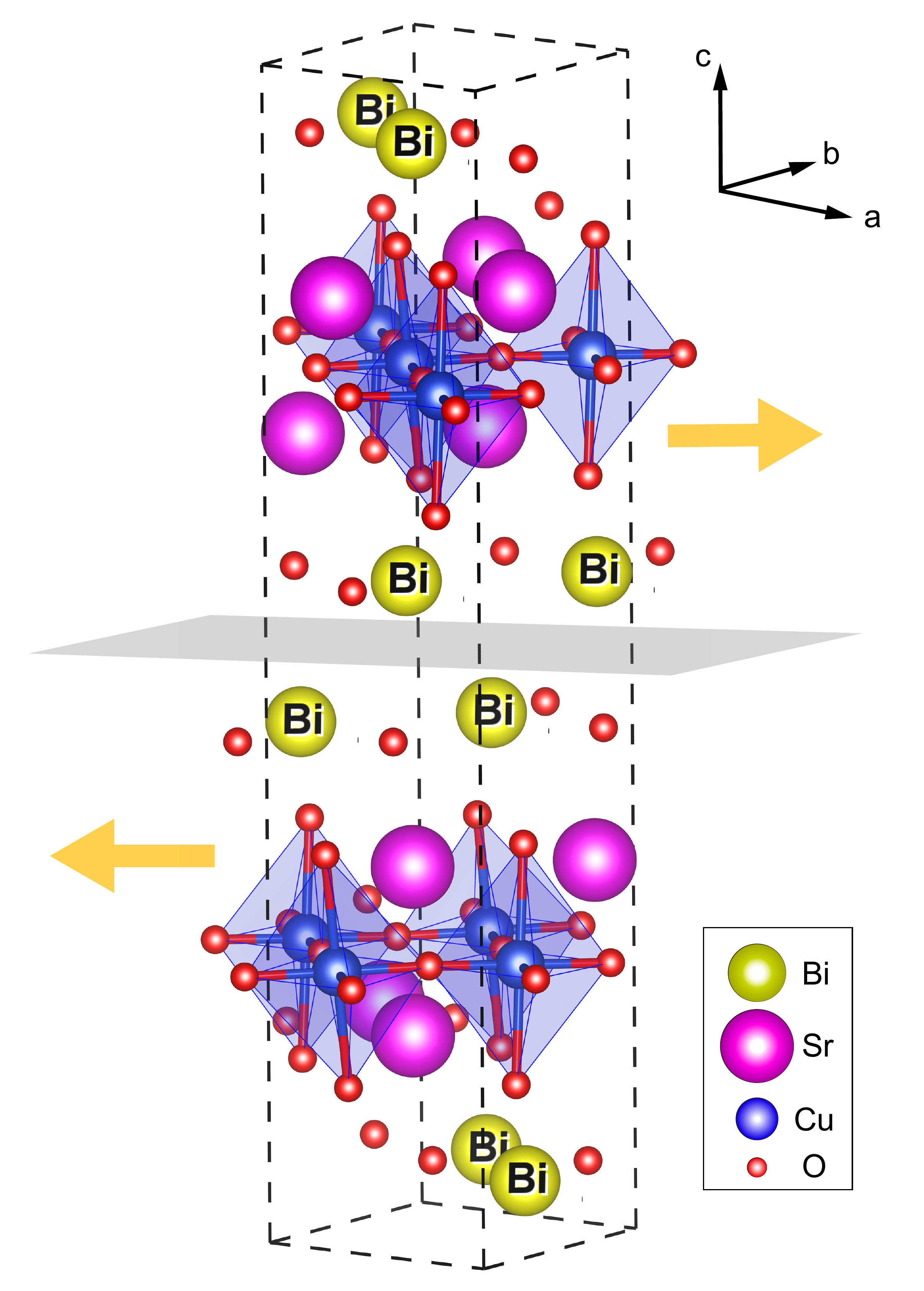}
    \caption{Bi2201 crystal structure. The unit cell parameters are $a \simeq 5.436~\textrm{\AA}$, $b \simeq 5.502~\textrm{\AA}$, $c \simeq 23.91~\textrm{\AA}$. The orange arrows indicate the shear motions within Bi-O layers discussed in the text.}
    \label{fig:crystal}
\end{figure}

\begin{figure*}[htbp]
    \centering
    \includegraphics[width=2\columnwidth,angle=0]{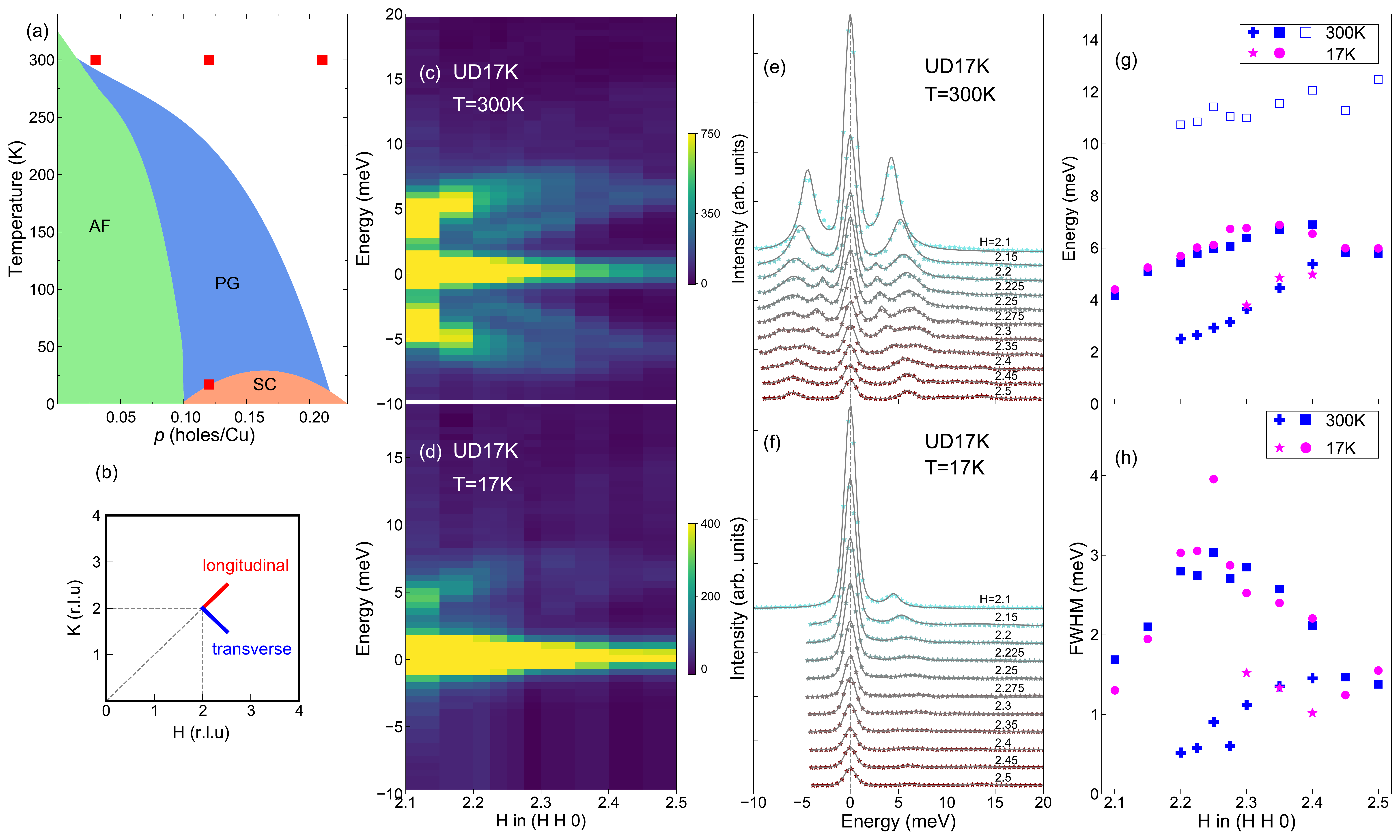}
    \caption{(a) Phase diagram of Bi2201, where the green shaded area indicates the antiferromagnetic (AF) region, the blue shaded area indicates pseudogap (PG) region and the orange region indicates superconductivity (SC) region. The red square markers indicate our measurements.
(b) Schematic plot of momentum space. The red line and blue line indicate our measurements along longitudinal direction and transverse direction, respectively. (c,d)
Temperature dependence for UD17K Bi2201. Low-energy longitudinal phonon dispersions along the (H H 0) direction at 300K and 17K, respectively. (e, f) Momentum dependence of the IXS spectra (markers) along the (H H 0) direction and the corresponding fittings (lines) overlapped on top. (g, h) Momentum evolution of fitting parameters for phonon dispersion and phonon intrinsic FWHM at 17K (magenta markers) and 300K (blue markers). Note that width values do not include the experimental energy resolution, which has been accounted by the resolution convolution. The various markers indicate different phonon modes with details in the text.}
    \label{fig:Temp}
\end{figure*}

We studied three doping levels of (Bi,Pb)$_2$(Sr,La)$_2$CuO$_{6+\delta}$ (Bi2201) as indicated in Fig.~\ref{fig:Temp}(a): antiferromagnetic (AF, p $\simeq$ 0.03), underdoped with $T_c$ = 17 K (UD17K, p $\simeq$ 0.12) and overdoped with $T_c$ = 11K (OD11K, p~$\simeq$~0.21). Underdoping was achieved via partial substitution of Sr with La to form Bi$_2$(Sr,La)$_2$CuO$_{6+\delta}$, while heavy overdoping was achieved via partial substitution of Bi with Pb to form (Bi,Pb)$_2$Sr$_2$CuO$_{6+\delta}$. The sample growth and characterization methods have been reported previously \cite{pengNC,zhaoCPL}. The IXS experiments were performed with HERIX spectrometer at Sector 30 of Advanced Photon Source at Argonne National Laboratory \cite{Said,Toellner}. The incident x-ray energy is 23.724 keV with an energy resolution of $\Delta$E $\sim$ 1.4 meV. The samples were glued to a sample holder inside a closed-cycle cryostat on a 4-circle goniometer. The Bi2201 crystal structure is shown in Fig.~\ref{fig:crystal} plotted using VESTA software \cite{VESTA}, where the adjacent Bi-O layers are held together by vdW forces. The reciprocal lattice units for Bi2201 throughout this paper are based on the orthorhombic cell convention \cite{chenTEM}. The IXS experiments were performed along both longitudinal direction and traverse direction around (2,2,0) as shown in Fig.~\ref{fig:Temp}(b) using transmission geometry. The lattice constants of Bi2201 determined from IXS measurements at 300 K are shown in Table~\ref{table:latcon}.

\begin{table}[t]
\caption{Comparison of experimental and calculated lattice constants (in \textrm{\AA}) for hole doped Bi2201 systems. Calculated constants are obtained with the rev-vdW-DF2 functionals, except for 0\%, which is obtained also by the PBE0 functional.} 
\centering 
\begin{tabular}{c c c c c c c c c c}
\hline \hline 
& & \multicolumn{3}{c}{Experiment} & & \multicolumn{4}{c}{Calculated} \\
\cmidrule{3-5} \cmidrule{7-9} 
& & AF & UD15K & OD11K && 0\% (PBE0) & 0\% & 12\% & 20\% \\
\hline
\midrule

$a$ && 5.436 & 5.4 & 5.4 &&    5.182 & 5.289 & 5.268 & 5.258 \\
$b$ && 5.502 & 5.445 & 5.308 && 5.347 & 5.362 & 5.375 & 5.366 \\
$c$ && 23.91 & 24.34 & 24.56 && 24.47 & 24.857 & 24.945 & 24.961 \\

\hline

\end{tabular}\label{table:latcon}
\end{table}

\begin{figure*}[htbp]
    \centering
    \includegraphics[width=2\columnwidth,angle=0]{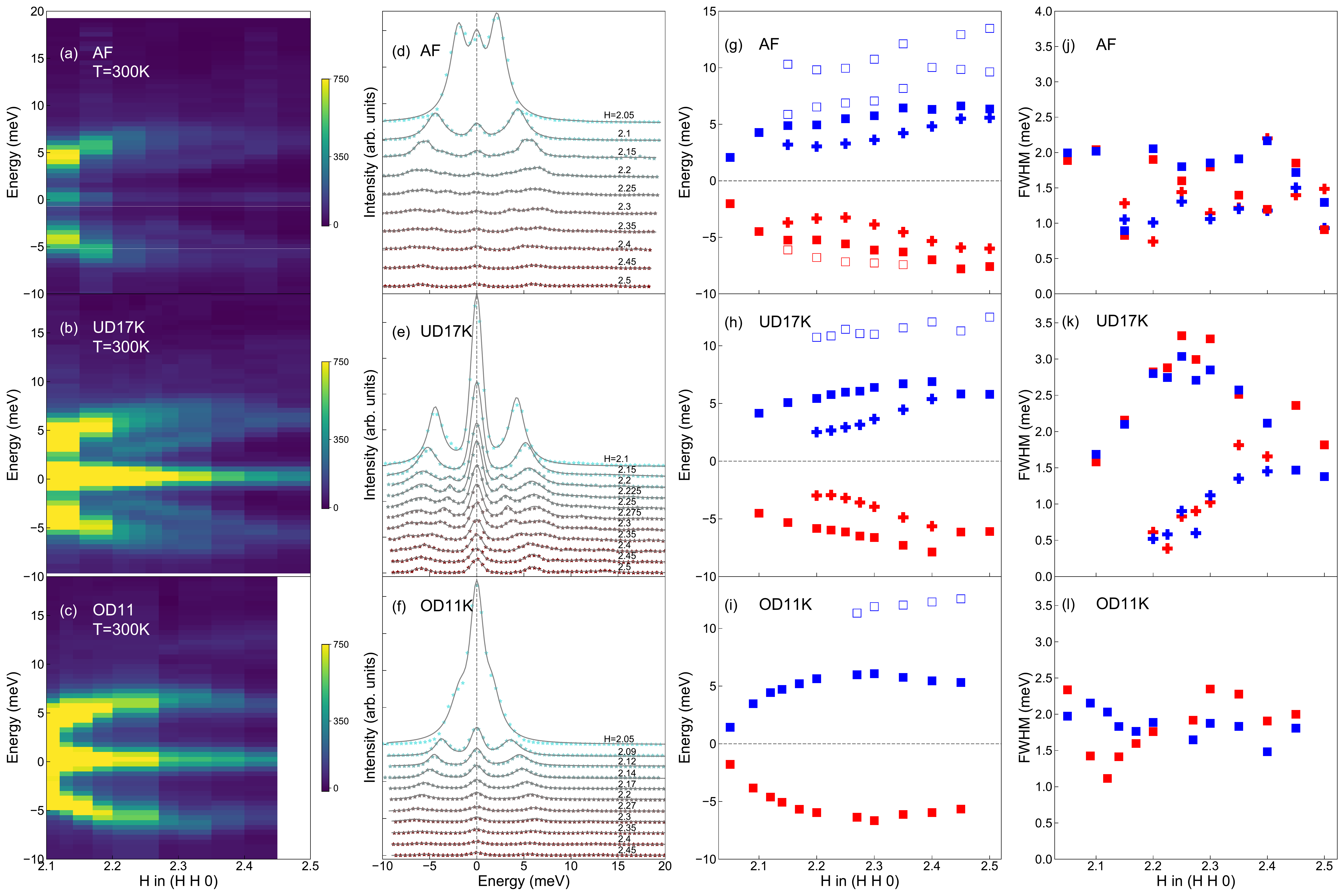}
    \caption{Doping dependence of longitudinal phonons for Bi2201 at 300K. Low-energy longitudinal phonon dispersions along the (H H 0) direction for AF (a), UD17K (b) and OD11K (c). (d, e, f) Momentum dependence of the IXS spectra (markers) along the (H H 0) direction and the corresponding fittings (lines) overlapped on top. Momentum evolution of fitting parameters for phonon dispersion (g, h, i) and phonon intrinsic FWHM (g, h, i) for AF, UD17K and OD11K, respectively. Note that width values do not include the experimental energy resolution, which has been accounted by the resolution convolution. The Stokes and anti-Stokes components are indicated as blue and red markers, respectively. }
    \label{fig:LA}
\end{figure*}

\begin{figure*}[htbp]
    \centering
    \includegraphics[width=2\columnwidth,angle=0]{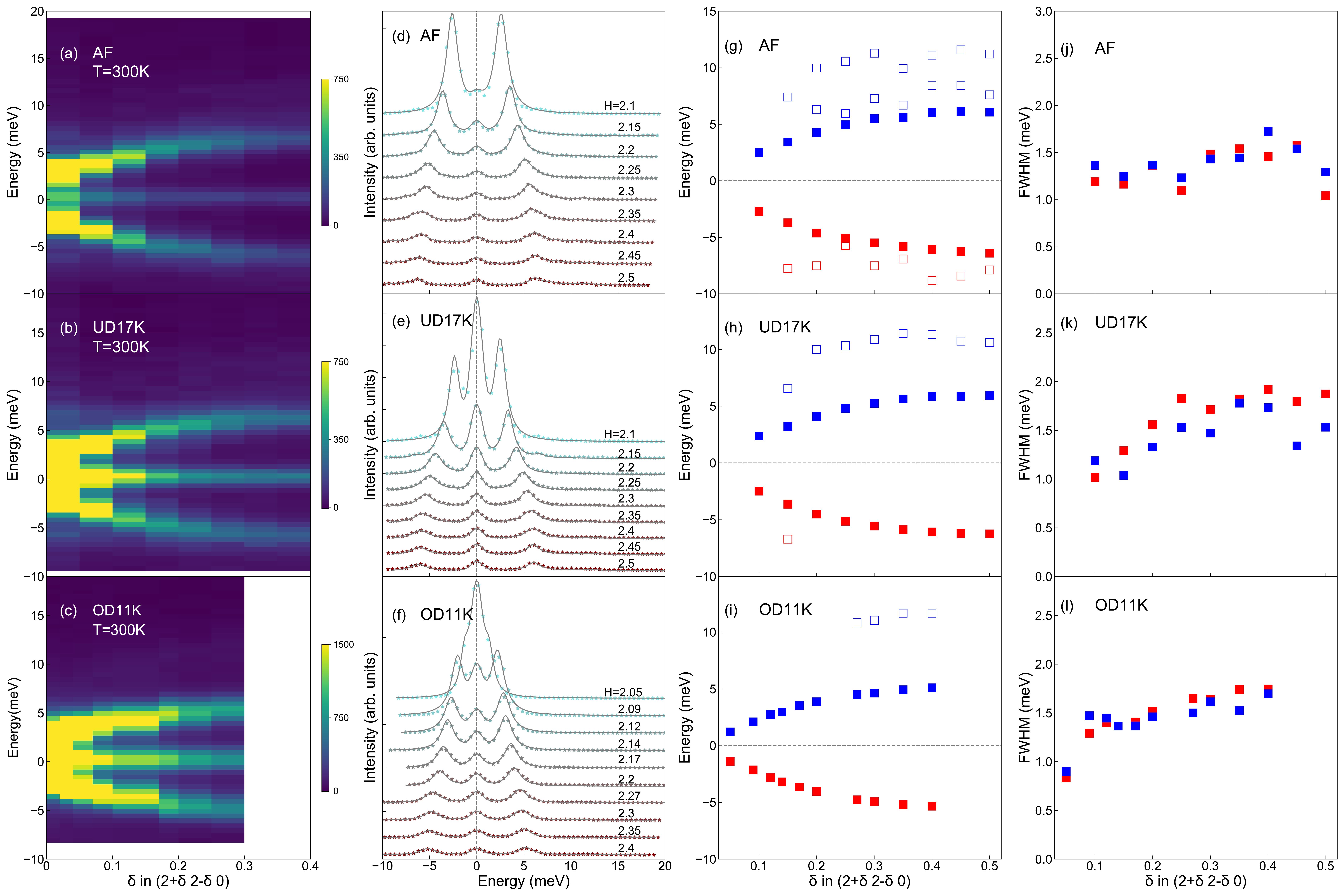}
    \caption{Doping dependence of transverse phonons for Bi2201 at 300K. Low-energy transverse phonon dispersions along the (2+$\delta$, 2-$\delta$, 0) direction for AF (a), UD17K (b) and OD11K (c). (d, e, f) Momentum dependence of the IXS spectra (markers) along the (2+$\delta$, 2-$\delta$, 0) direction and the corresponding fittings (lines) overlapped on top. Momentum evolution of fitting parameters for phonon dispersion (g, h, i) and phonon intrinsic FWHM (g, h, i) for AF, UD17K and OD11K, respectively. Note that width values do not include the experimental energy resolution, which has been accounted by the resolution convolution. The Stokes and anti-Stokes components are indicated as blue and red markers, respectively. }
    \label{fig:TA}
\end{figure*}

\subsection{Theory}
Low energy vibrational properties of Bi2201 associated with the shear motions of the Bi-O planes were analyzed using phonon calculations based on ground state DFT. As noted previously by a number of authors, the observed geometry of Bi2201 crystals exhibits an incommensurate superstructure and is not amenable to a description based on small periodically repeating unit cells \cite{HEINRICH1994133,PhysRevB.51.3128}. In particular the superstructure involves incommensurate modulations of the Bi-O sublattice \textcolor{black}{which manifest as a variety of different Bi-O bonding configurations between and within the sublattices \cite{Bi2212slab}}. For analyzing the electronic properties of Bi2201, such as the Fermi surface \cite{PhysRevB.51.3128}, theoretical methods therefore employed an approximate periodic structure based on the $\mathrm{\sqrt{2} \times \sqrt{2}}$ orthorhombic (ORTH) cell determined by Torardi et al \cite{Torardi_cell}, which is similar to the schematic structure shown in Fig~\ref{fig:crystal}. Simulations of phonon properties in this study are also based on periodic supercells constructed from the approximate ORTH structure. The optimized geometries of such small periodic cells however do not correspond to the global minimum structure of Bi2201 (which is incommensurate) but instead represent either nearby local minima or alternately saddle points on the high dimensional potential energy surface (PES). In order to extract a consistent and useful qualitative description of Bi2201 phonons from periodic cell models we carried out different sets of simulations: First, in order to visualize atomic displacements associated with both zone center and zone boundary normal modes of interest, we carried out phonon calculations using the CRSYTAL17 package~\cite{dovesi2005crystal,dovesi2009crystal09}. These calculations also employed both hybrid and semi-local exchange-correlation (XC) functionals to assess the robustness of the description with respect to choice of XC functional. Next, in order to map out phonon dispersions across the Brillouin zone and make contact with experimental IXS spectra, we performed frozen phonon simulations with the VASP~\cite{Kresse_1995} and phonopy~\cite{Parlinski,phonopy} codes employing dispersion correct vdW DFT functionals~\cite{Hamadavdw}. Numerical parameters characterizing both sets of simulations are listed in the following. 

Calculations with the CRYSTAL17 package \cite{dovesi2005crystal,dovesi2009crystal09} carried out geometry optimizations on undoped Bi2201 with the ORTH structure using both semilocal PBE~\cite{Perdew1996}   
and non-local hybrid PBE0~\cite{Adamo1999} XC functionals. The hybrid functional, was previously calibrated versus QMC results for the cuprate spin, charge, and phonon behavior \cite{wagner2014effect}. The pseudopotentials by Burkatzki, Filippi, and Dolg \cite{burkatzki2007energy,burkatzki2008energy} were used to remove core electrons.
The wave functions were represented by the triple-zeta basis set included with those potentials.
A k-point mesh of $4 \times 4 \times 4$ was used to integrate over the Brillouin zone according to a Pack-Monkhorst sampling. The optimized lattice parameters for the AF configuration of the ORTH structure are given by $a = 5.182~\textrm{\AA}$, $b = 5.347~\textrm{\AA}$, and $c = 24.47~\textrm{\AA}$. Phonon frequency calculations were performed using a frozen phonon approach at $\vec{q} = (0, 0, 0)$ and $(1, 1, 0)$. 

For phonon band structure simulations with dispersion corrected functionals, DFT with projector-augmented wave pseudopotentials was employed as implemented in the Vienna \textit{ab initio} simulation package (VASP) code \cite{kresse1993ab,kresse1994ab,kresse1996efficiency,kresse1996efficient}. Van der Waals dispersions between Bi-O layers are accounted for by utilizing the rev-vdW-DF2 functional \cite{Hamadavdw} which has been shown to be relatively accurate in calculating interlayer and intralayer lattice constants in layered solids \cite{TranBenchmark}. The on-site Coulomb repulsion U was set to 6~eV for the Cu 3d orbitals in the system. In consideration of the incommensurate Bi-O modulations in the real system~\cite{Torardi_cell}, the vdW DFT simulations investigated two antiferromagnetic configurations based on the ORTH structure with modified Bi-O bonding environments as shown in the Appendix Fig.~\ref{fig:two_structures}. The first structure (termed "XY") in Fig.~\ref{fig:two_structures}(a) has been modified to exhibit Bi-O intralayer dimerization along both the $a$- and $b$- lattice directions instead of only along the $b$-lattice direction present in the ORTH structure. The second considered structure in Fig.~\ref{fig:two_structures}(b) follows from the results of Nokelainen et al. who used the SCAN functional to determine a Bi-O bonding configuration of stacking zigzag chains in Bi2212 \cite{Bansil_zz}. Both variations were modified starting from the ORTH structure of Torardi et al~\cite{Torardi_cell} obtained from the ICSD \cite{ICSD}.

 Geometry optimizations of the initial structure were performed by relaxing all degrees of freedom until the Hellman-Feynman forces were less than 0.001~eV/$\textrm{\AA}$ with a $6 \times 6 \times 1$ $\Gamma$-centered k-point sampling mesh and a plane wave cut-off energy of 520 eV. Table \ref{table:latcon} summarizes the final relaxed lattice constants for the investigated systems using these parameters. Phonon dispersions were calculated using the Phonopy package \cite{phonopy} with a $2 \times 2 \times 1$ supercell (176 total atoms) and a $3 \times 3 \times 1$ k-point mesh.  To qualitatively reproduce the trends seen experimentally with doping dependence of the low energy phonons, we consider three different hole doping levels \textcolor{black}{for the XY structure} -- 0\% (undoped), 12\%, and 20\% doping -- through the virtual crystal approximation \cite{Kronik_VCA}. In this way, Pb doping is effectively introduced into the system by modifying the Bi nuclear charge to be the average of the Bi and Pb nuclear charges. Similarly three doping levels of 6$\%$, 10$\%$ and 14$\%$ were analyzed for the zigzag bonding structure. For each motif (XY or Zigzag), the doping levels were chosen from a doping range in which the optimized structures present a minimum number of unstable phonon modes. We also note that in the range of investigated dopings, the AF configuration remained magnetically stable relative to a ferromagnetic one. An analysis of other modulated magnetic orderings such as stripes which require larger supercells is beyond the scope of this study. Simulated IXS spectra were made using the SNAXS software package \cite{SNAXS}. Structures were visualized with VESTA \cite{VESTA}.

\section{Experimental results}
\subsection{Observation of van der Waals modes}

We performed our IXS experiment around the orthorhombic (2, 2, 0) peak since it is one of the strongest accessible Bragg reflections. We have measured UD17K at 300K and 17K (T$_c$) and the momentum-energy intensity color maps are shown in Figs.~\ref{fig:Temp}(c) and (d), respectively. Here, the positive energy corresponds to photon energy loss, while the negative energy corresponds to photon energy gain. There is no clear charge ordering peak around (2.25, 2.25, 0) emerging in the elastic peak as observed by using soft x-ray resonant scattering at the Cu $L_3$-edge \cite{PengUDBi2201,Damascelliscience}. This is consistent with the previous IXS results of Bi2212 \cite{HePRB} but different from the quasi-elastic peak found in YBCO \cite{TaconYBCO}. This may arise from the comparatively weaker and shorter-ranged charge modulations in Bi-based cuprates. We also observed symmetric Stokes and anti-Stokes components at 300K due to phonon creation and annihilation, while only the Stokes phonons remain at 17K due to the Bose factor.

The IXS spectra at varying reciprocal momenta for 300K and 17K are shown in Figs.~\ref{fig:Temp}(e) and (f), respectively. The spectra are shifted vertically for clarity. The spectra are composed of the elastic component centred at zero energy and the inelastic components due to Stokes and anti-Stokes scattering from phonons. The elastic component is fitted with an intensity-scaled resolution function. The phonon excitations were fitted to the resolution-convolved Lorentz functions with the Bose factor correction. The fittings for one selected IXS spectrum are shown in Fig.~A1 in the Appendix. The results of these fits for phonon dispersions and the intrinsic full-width at half-maximun (FWHM) are displayed in Figs.~\ref{fig:Temp}(g) and (h). We can clearly identify three phonon modes from the phonon dispersions. The two higher energy modes with energy ranges $\sim$10-12 meV and $\sim$4-7 meV are similar to those observed in bilayer Bi2212, which were assigned to the longitudinal acoustic mode and one low-energy optical mode \cite{HePRB}.

The lowest-energy phonon has an energy range from 2 meV to 5 meV that is too low for an acoustic phonon. Moreover, it does not go to zero at (2,2,0) considering its curvature. This mode is not observed in other cuprates, such as YBCO \cite{TaconYBCO} and LBCO \cite{miao2018incommensurate}. On the other hand, this is similar to the very low-energy vdW modes ($\sim$2-5 meV) observed in a serious of layered materials, such as few-layer graphene \cite{ShearModesGraphene} and TMD materials \cite{ShearModes}. We demonstrate below that this ultra-low energy phonon is the vdW mode of Bi2201 resulting from the interlayer vdW restoring force, which induces the in-plane shear motions between Bi-O layers with half unit-cell displacing together. 

Moreover, we did not observe any evidence of phonon softening at the charge order wave vector (2.25, 2.25, 0) as shown in Fig.~\ref{fig:Temp}(g), which is different from those found in YBCO and LBCO \cite{TaconYBCO,miao2018incommensurate}. On the other hand, the FWHMs for those two low-energy branches with prominent IXS intensity are shown in Fig.~\ref{fig:Temp}(h). We have observed a width broadening of $\sim$3 meV for the low-energy optical mode around (2.25, 2.25, 0), which has been also observed at similar momentum in Bi2212 \cite{HePRB}. It has been proposed that the broadening of phonon width is due to the simultaneous measurement of multiple phonons close in energy to each other, thus the amplitude of the eigenvector changes throughout the zone and shows up prominently at (2.25, 2.25, 0) \cite{HePRB}. Interestingly, this broadening effect is not found in Pb-substitution underdoped Bi2201 samples where Pb would suppress the well-known structural incommensurate supermodulation \cite{BonnoitPhD}. This suggests that the phonon dispersions and widths are very sensitive to the details of the crystal structure rather than the presence of charge order. Note that the FWHM of the vdW mode is only $\sim$1 meV after deconvolving the energy resolution. Both phonon dispersions and FWHMs do not show any obvious change at 17K and 300K. 

\subsection{Doping dependence}

We now turn to the doping dependence of the low-energy phonons in Bi2201. First, we studied the longitudinal phonons (LA) of Bi2201 at 300K, as shown in Fig.~\ref{fig:LA}. The Stokes and anti-Stokes phonons show very symmetric behaviors, which can be noticed clearly in Figs.~\ref{fig:LA}(a-c). One additional phonon branch ($\sim$5-10 meV) can be identified in the AF sample (Fig.~\ref{fig:LA}(g)) compared to UD17K (Fig.~\ref{fig:LA}(h)). The low-energy vdW mode is obvious in AF and UD17K, whereas its intensity fades in OD11K. Moreover, the energy of the vdW mode gets lower in UD17K than in AF sample. We notice that the $c$-axis lattice parameter increases with doping (Table~\ref{table:latcon}). This indicates the interlayer coupling is influenced by the $c$-axis spacing, which is confirmed by the calculations shown below. We expect the vdW mode would persist in one unit-cell Bi-based cuprates but having a smaller energy, which would be similar to the layers dependence of shear modes reported in multilayer graphene \cite{ShearModesGraphene}, MoS$_2$ and WSe$_2$ \cite{ShearModes}. This is anticipated by a decrease of restoring force with decreasing number of layers \cite{ShearModesGraphene}. The phonon FWHMs for the most intense branch and vdW mode are shown in Figs.~\ref{fig:LA}(j-l). The FWHMs from both Stokes (blue markers) and anti-Stokes (red markers) components agree well with each other. Interestingly, the phonon broadening at (2.25, 2.25, 0) r.l.u. for the low-energy optical mode only exists in UD17K and disappears in both AF and OD11K samples, confirming this broadening is very sensitive to structure.

We also display the transverse cuts taken along (2+$\delta$, 2-$\delta$, 0) direction for three dopings of Bi2201 at 300K, which is shown in Fig.~\ref{fig:TA}. The transverse branches appear at lower energies ($\sim$2 meV) than the low-energy longitudinal branches. The phonon FWHMs for the most intensive branch is shown in Figs.~\ref{fig:TA}(j-l). Clearly, there is no phonon width broadening for the transverse branches as shown in Figs.~\ref{fig:TA}(j, k, f). Moreover, the vdW mode is absent for all three dopings for transverse cuts (Figs.~\ref{fig:TA}(g, h, i)), which may relate to the IXS cross sections. According to the polarization dependence of the IXS cross section from phonons, which is proportional to $(\vec{Q}.\vec{\epsilon})^2$ where $\vec{\epsilon}$ is the phonon eigenvector \cite{IXSreview}, this suggests the vdW mode has no transverse component.

\begin{figure}[htbp]
    \includegraphics{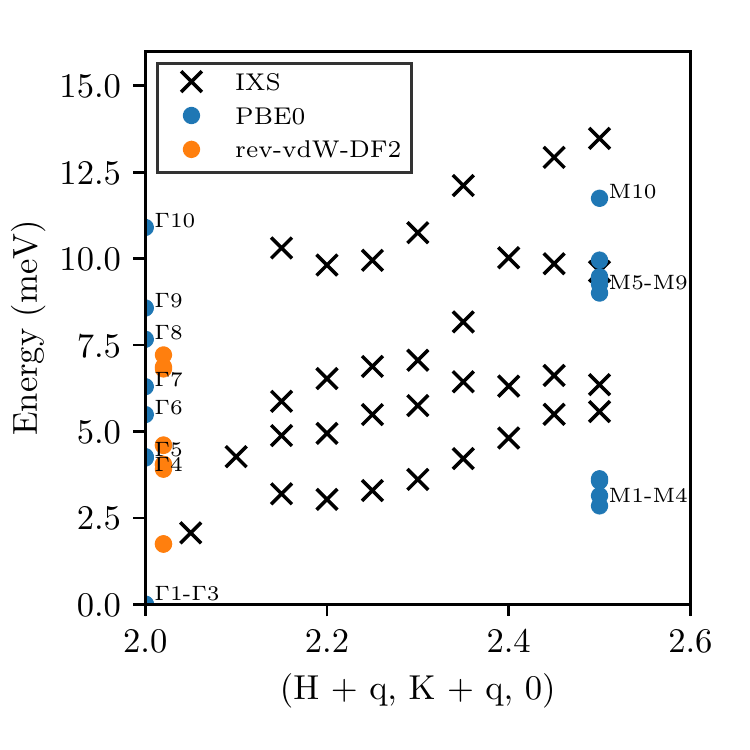}
    \caption{Theoretical and experimental phonon frequencies for Bi2201. The theoretical phonons were computed using DFT in the PBE0 approximation. We associate the M1 to M4 modes with the modes seen experimentally at the Brillouin zone boundary, and the $\Gamma4$ and $\Gamma5$ modes to the optical modes at $\Gamma$.
    } 
    \label{fig:disp_pbe0}
\end{figure}

\begin{figure*}[htbp]
    \centering
    \includegraphics{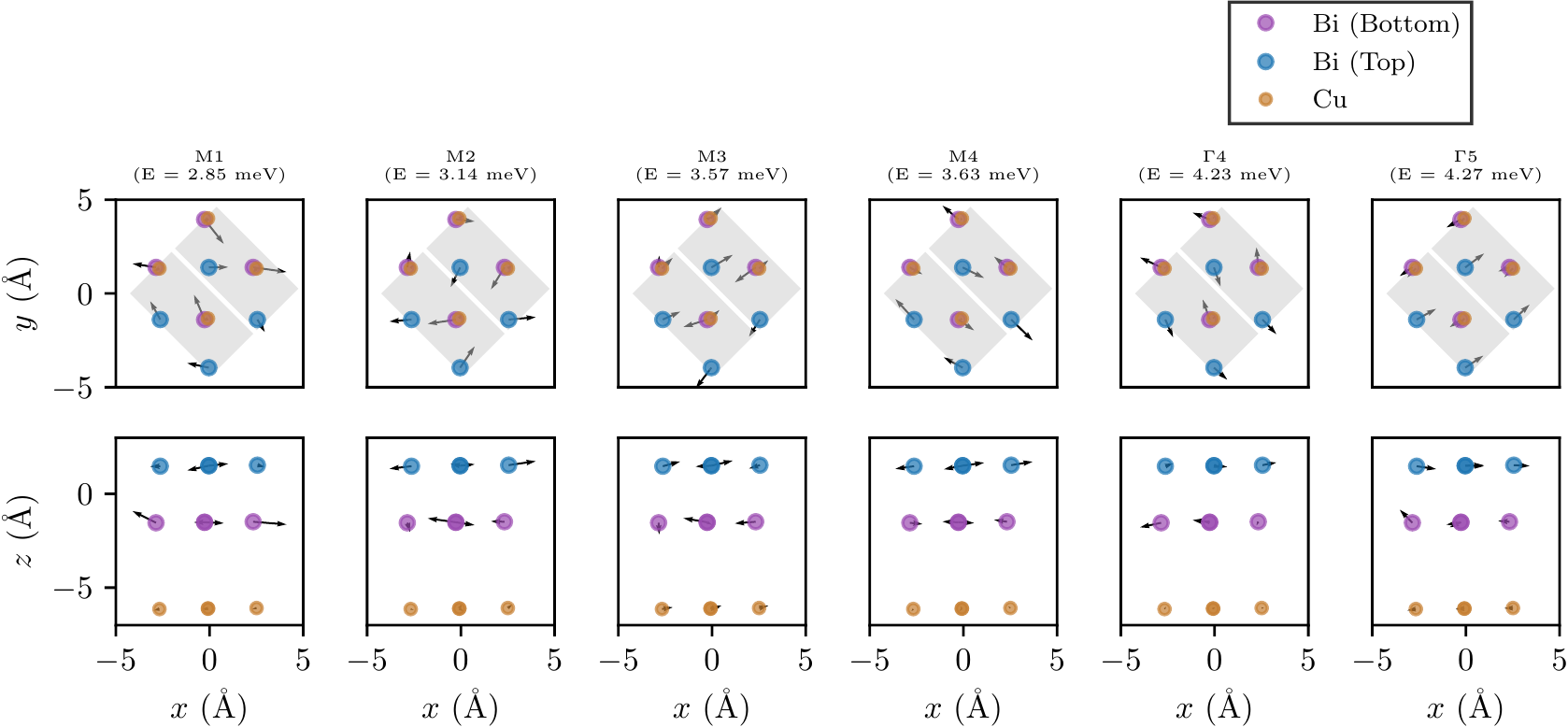}
    \caption{Phonon eigenvectors from DFT (PBE0) calculations. Oxygen atoms are not included in the plots to clearly visualize the displacement of Bi atoms. Each column displays a phonon mode, with the top layer showing the top-down view and the bottom layer showing the side view.}
    \label{fig:eigvec_pbe0}
\end{figure*}

\section{Theory results}

The experimental observations presented in the previous section can be understood qualitatively on the basis of first-principles phonon simulations. A quantitative comparison is however hindered as the superstructure modulation of the Bi-O sublattice is not accounted for in the simulations. Regardless of the methodology employed to calculate the phonons, we find that \textcolor{black}{(with exception for a narrow regime of doping levels on the Bi-O zigzag bonding structure)}, models based on the approximate ORTH periodic unit cell predict unstable modes at wave vectors away from the zone boundary. Nevertheless, the modes at the zone-center as well as band dispersions along certain high-symmetry directions in the BZ are well behaved and are analyzed here. 

First, we analyze the character of the low energy vdW modes using the PBE0 functional. 
The calculated phonon frequencies are reported in Table~\ref{tab:disp} and plotted along with the IXS and rev-vdW-DF2 results in Fig.~\ref{fig:disp_pbe0}. The labels $\Gamma$ and M represent phonon modes at $\vec{q} = (2, 2, 0)$ and $(2.5, 2.5, 0)$, respectively.
The figure shows low-energy optical and acoustic branches mixing together into four nearly degenerate phonons at the Brillouin zone boundary. 
The frequencies from PBE0 are in good agreement with the rev-vdW-DF2, except for the two additional low-energy bands at $1.74~\rm{meV}$ and $1.75~\rm{meV}$, which are discussed in the following paragraph.
The phonon eigenvectors are displayed in Fig.~\ref{fig:eigvec_pbe0}, with the top row being the top view and the bottom row being the side view.
From the diagram, the M1 and M2 modes are associated with rotation of the Bi planes, while the M3, M4, $\Gamma 4$, and $\Gamma 5$ modes are associated with plane shearing.
All low-energy modes are due to the the weakly bound cleavage plane.
From the calculations, we ascertain that vdW modes are primarily associated with a- and b-direction motions of the Bi atoms in adjacent Bi-O layers, explaining why the vdW functional obtains different frequencies from the hybrid functional. 

Based on structure optimizations using the vDW functional, we find that although the Bi-O zigzag bonding variation predicts a lower ground state energy compared to the XY variation (on the order of 0.1 eV per unit cell), both structures predict unstable phonon modes away from the zone boundary when undoped. Such instabilities cannot be eliminated by considering energy lowering distortions that fit within small periodic cells. However in the Bi-O zigzag structure within the range of  6\% -- 14\% hole doping, no unstable modes are seen suggesting that this structure is at least a local minimum in the PES over this doping range. Further, our modified ORTH XY structure experiences an antiferromagnetic to ferromagnetic phase transition at 26\% hole doping. This is in good agreement with the two-dimensional ferromagnetic fluctuations observed in overdoped Bi2201, where the magnetic ground state changes from antiferromagnetic to ferromagnetic with increasing doping \cite{FerromagneticBi2201}. However, we simplify our qualitative analysis by including only the antiferromagnetic configurations up to 20\% hole doping. 

\begin{figure}[htbp]
    \includegraphics[width=\columnwidth,angle=0]{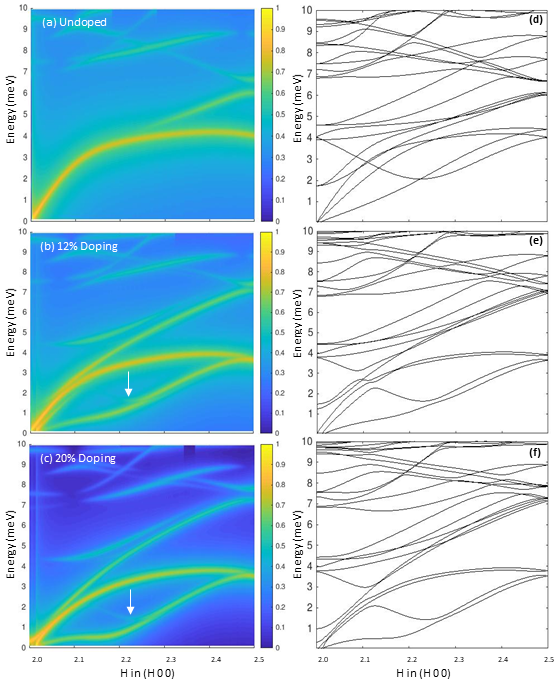}
    \caption{Theoretical doping dependence of phonons for Bi2201 using the rev-vdW-DF2 functional \textcolor{black}{on the XY structure}. Simulated IXS spectra along the (H 0 0) direction for 0\% doping (a), 12\% doping (b), and 20\% doping (c). (d, e, f) Calculated phonon dispersions for 0\%, 12\%, and 20\% doping, respectively. Intensities have been normalized and plotted on a logarithmic scale for clarity. All systems are in the  antiferromagnetic state.}
    \label{fig:Sim_IXS}
\end{figure}

Figure \ref{fig:Sim_IXS} shows the simulated IXS and calculated dispersions of three hole doping levels considered within the frozen-phonon method and the rev-vdW-DF2 functional on the modified XY ORTH structure.~Both the XY and zigzag configurations predict the vdW phonon modes measured experimentally but the latter structure exhibits very low simulated IXS intensity (phonon dispersions and simulated IXS are shown in the Appendix). Because the incommensurate modulations in the true Bi2201 system allow for numerous potential Bi-O bonding configurations to co-exist \cite{Bi2212slab}, we hypothesize that the $a-$ and $b-$direction Bi-O dimerization motif similar to the one present in our XY structure is responsible for the experimentally measured vdW IXS intensity. Therefore for a comparison with experimental IXS results, Figure \ref{fig:Sim_IXS} is presented with the XY structure. Additionally we note that the dispersions and intensities for the XY structure are presented along the (200) high symmetry direction of Bi2201 whereas experimental observations are made along the (220) direction. As before, this choice is made to reveal qualitative trends while avoiding complications arising from the incommensurate superstructure that is not modeled. 

From the calculated volume studies of our doped systems (Table~\ref{table:latcon}) we see a similar trend with experimental observations, namely that the $c$-lattice constant increases with increasing doping level (from $c=24.857~\textrm{\AA}$ in the undoped system to about $c=24.961~\textrm{\AA}$ in the maximally doped system). As alluded to before, we expect this increase of the $c$-axis dimension to weaken the interlayer vdW coupling between Bi-O layers in the Bi2201 cell, causing a reduction in vdW phonon mode energy. Indeed, the modes marked by white arrows in Fig.~\ref{fig:Sim_IXS}(b, c) show this softening  between 12\% and 20\% doped systems. We note that the same mode is not IXS active in the undoped system (Fig.~\ref{fig:Sim_IXS}(a)). The hybridization between different modes is modulated across the doping regime causing the relative intensities of modes to vary. \textcolor{black}{A similar volume study preformed on the Bi-O zigzag structure (results shown in the Appendix) also predicts a $c$-lattice expansion (from $c=24.5535~\textrm{\AA}$ in the minimally doped system to about $c=24.5571~\textrm{\AA}$ in the maximally doped system). The lesser degree of expansion compared to experimental observations supports our hypothesis that the experimental system exhibits more $a$- and $b$-direction Bi-O dimerization motifs than the zigzag chain bonding.}

To summarize, our theoretical analysis indicates that the low energy phonon modes in Bi2201 are of vdW character involving shear motions along the cleavage plane. Furthermore, these vdW modes soften under hole doping in conjunction with the $c$-axis lattice expansion.

\section{Conclusion}

With high-resolution IXS we have measured the low-energy phonon dispersions in the high-$T_c$ compound Bi2201 and uncovered the vdW mode in cuprates for the first time. The vdW phonons are in good agreement with first-principles DFT calculations. Our result demonstrates the vdW mode is generic for layered materials with interlayer vdW interactions even for systems as extremely chemically complex as doped copper oxides. This is important for the out-of-plane alignments, such as interlayer twist and vertical displacements. By utilizing the vdW interactions one can exploit Bi-based cuprates to engineer novel heterostructures, similar to graphene and transition metal dichalcogenides. For example, it has been proposed that high-temperature topological superconductivity can be realized in twisted Bi2212 by mechanically exfoliating and the control of the twisted angles in the vicinity of 45$^0$ \cite{twistedBi2212}.

\section{Acknowledgement}

This work was supported by the QSQM, an Energy Frontier Research Center funded by the U.S. Department of Energy (DOE), Office of Science, Basic Energy Sciences (BES), under Award $\#$DE-SC0021238. P.A. gratefully acknowledges support from the EPiQS program of the Gordon and Betty Moore Foundation, grant GBMF9452. This research used resources of the Advanced Photon Source, a U.S. Department of Energy (DOE) Office of Science User Facility operated for the DOE Office of Science by Argonne National Laboratory under Contract No. DE-AC02-06CH11357.
This material is based upon work (L.K.W) supported by the U.S. Department of Energy, Office of Science, Office of Basic Energy Sciences, Computational Materials Sciences program under Award Number DE-SC-0020177. L.Z. and X.J.Z thank the financial support from the National Natural Science Foundation of China (Grant no. 11888101), the National Key Research and Development Program of China (Grant no. 2016YFA0300300) and the Strategic Priority Research Program (B) of the Chinese Academy of Sciences (Grant no. XDB25000000). Simulation work by I.B, T.P.D and C.D.P was supported by the U.S. Department of Energy, Office of Basic Energy Sciences, Division of Materials Sciences and Engineering, under Contract No. DE-AC02-76SF00515 through TIMES at SLAC. Theory simulations used resources of the National Energy Research Scientific Computing Center (NERSC), a U.S. Department of Energy Office of Science User Facility operated under Contract No. DE-AC02-05CH11231.

\newcommand{\beginsupplement}{
        \setcounter{table}{0}
        \renewcommand{\thetable}{A\arabic{table}}
        \setcounter{figure}{0}
        \renewcommand{\thefigure}{A\arabic{figure}}
     }
     
\beginsupplement

\section{Appendix}

\subsection{IXS data fitting}

\begin{figure}
    \includegraphics[width=1\columnwidth,angle=0]{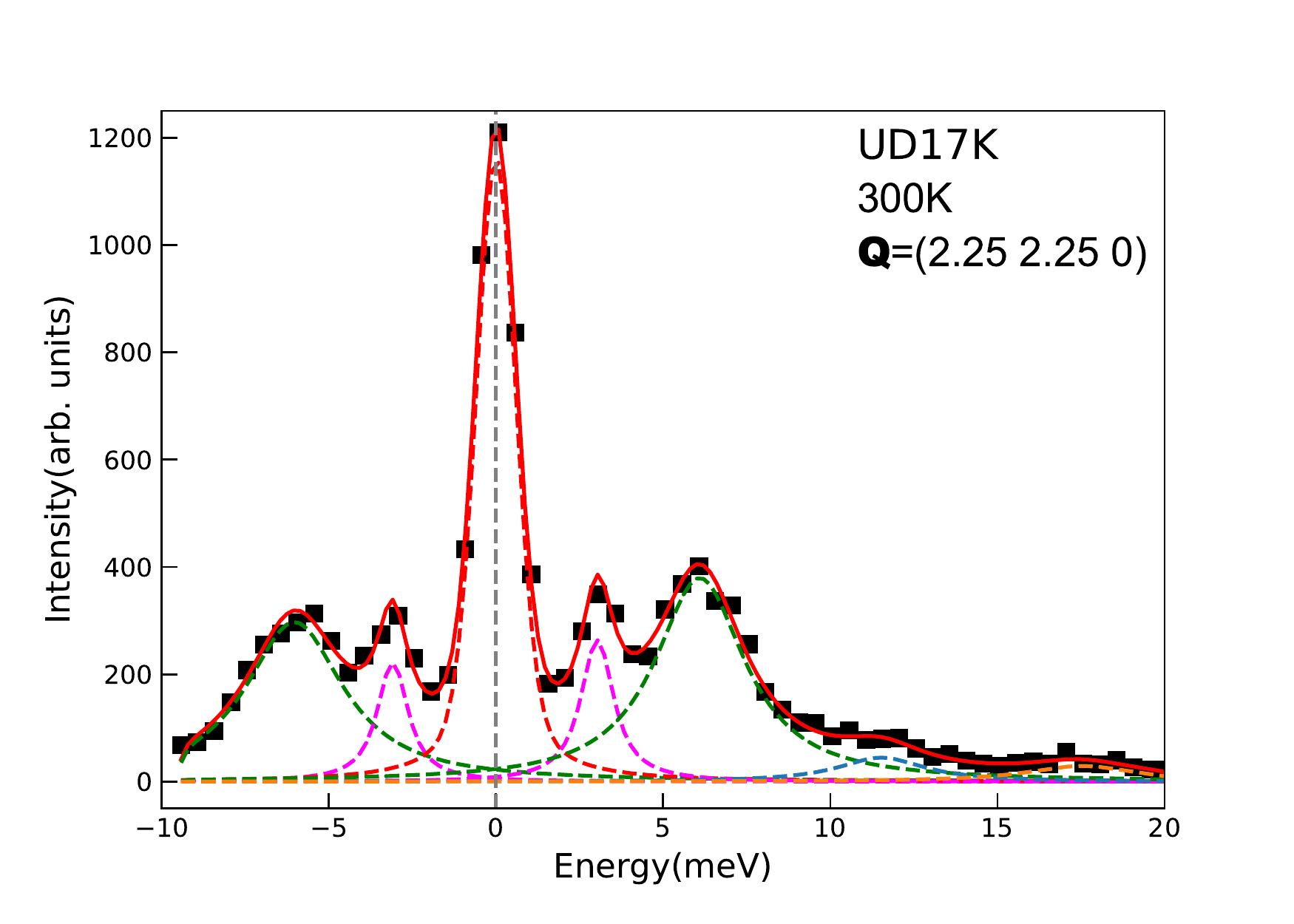}
    \caption{Example fit of one selected IXS spectrum of UD17K taken at 300K at $\boldsymbol{Q}$=(2.25,2.25,0). The dashed red line represents the resolution function profile with 1.4-meV FWHM. The dashed magenta and green lines represent the results of fits to the inelastic intensity. The solid red line is the sum of these contributions. } 
    \label{fig:fitting}
\end{figure}

The experimental data were fitted using Lorentz functions, convoluted with the experimental resolution, as shown in Fig.~\ref{fig:fitting}. The position and linewidth of the phonons reported in the paper are deconvoluted from the resolution.

\subsection{Phonon calculations for Bi-O zigzag ORTH structure}
\begin{figure}
    \includegraphics[width=1\columnwidth,angle=0]{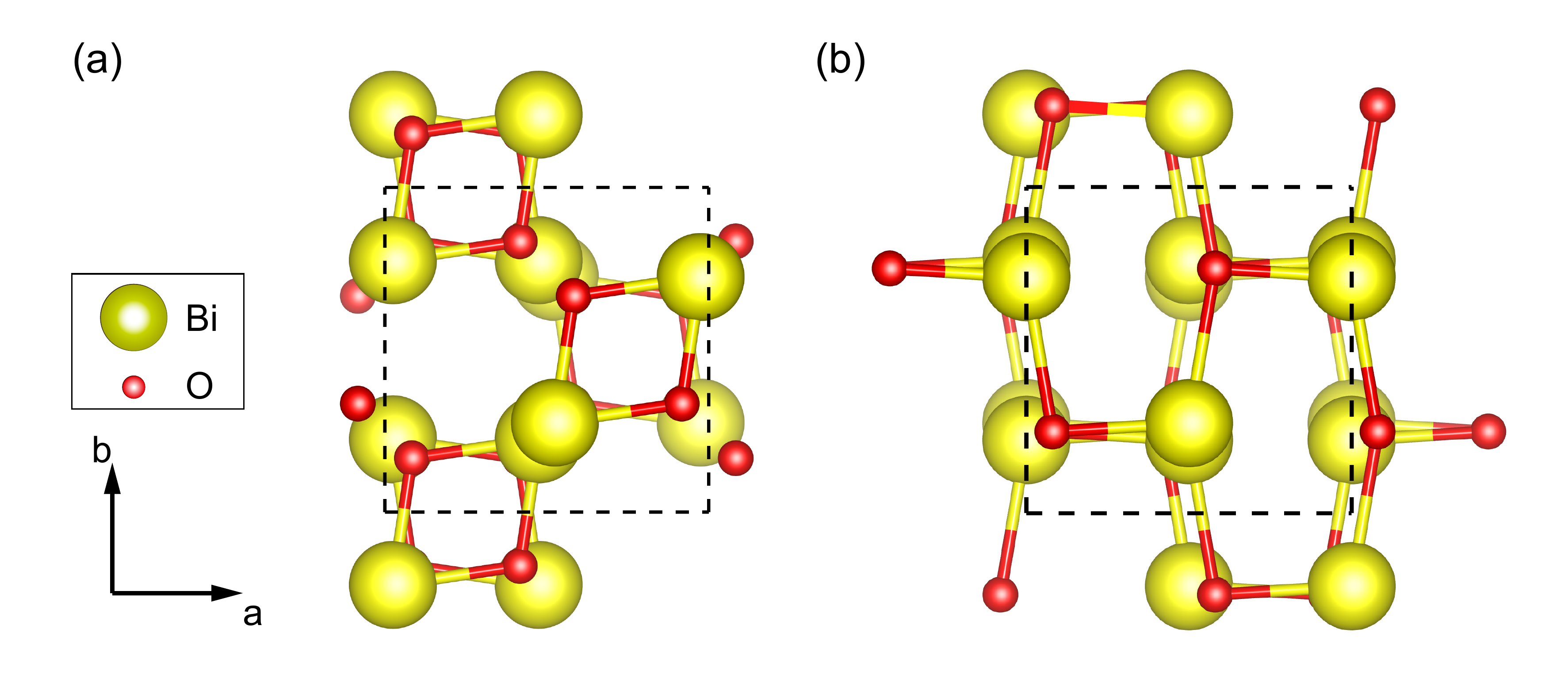}
    \caption{Top view of the Bi-O layers for XY structure (a) and zigzag structure (b) of Bi2201.} 
    \label{fig:two_structures}
\end{figure}

\begin{figure}[htbp]
    \includegraphics[width=\columnwidth,angle=0]{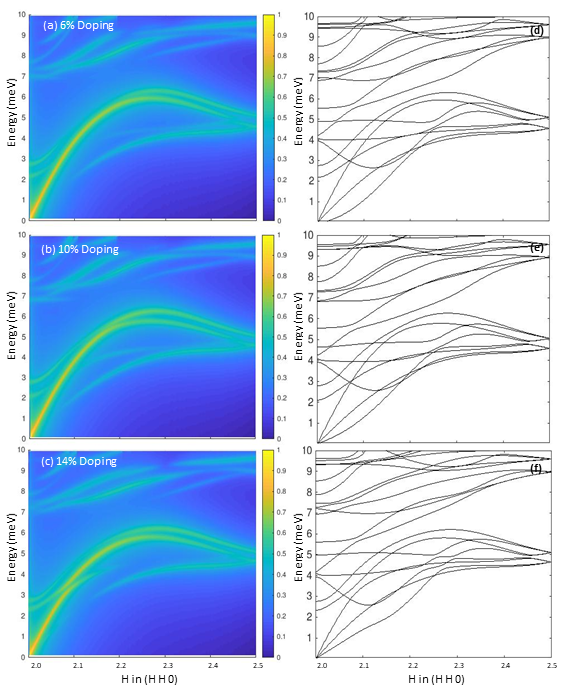}
    \caption{Theoretical doping dependence of phonons for Bi2201 using the rev-vdW-DF2 functional on the Bi-O zigzag structure. Simulated IXS spectra along the (H H 0) direction for 6\% doping (a), 10\% doping (b), and 14\% doping (c). (d, e, f) Calculated phonon dispersions for 6\%, 10\%, and 14\% doping, respectively. Intensities have been normalized and plotted on a logarithmic scale for clarity. All systems are in the stable antiferromagnetic state and predicted no unstable phonon modes.}
    \label{fig:Sim_IXS_ZZ}
\end{figure}

\textcolor{black}{Figure \ref{fig:Sim_IXS_ZZ} shows the simulated IXS and calculated dispersions of three hole doping levels using the Bi-O zigzag bonding structure as shown in Fig.~\ref{fig:two_structures}(b). The hole doping levels were chosen such that no phonon mode instabilities were predicted within the presented doping range, thus allowing for the IXS simulations to be preformed along the experimentally observed (220) direction. We note that the vdW modes have low intensity in this structure and we hypothesize the dimerization motif represented by our XY structure is responsible for the IXS intensity of the vdW modes. This is further supported by the reduced $c$-lattice expansion in the zigzag bond configuration (Table~\ref{table:latcon_ZZ}).}

\begin{table}[t]
\caption{Comparison of experimental and calculated lattice constants (in \textrm{\AA}) for hole doped Bi2201 systems. Calculated constants are obtained with the rev-vdW-DF2 functional on the Bi-O zigzag structure.} 
\centering 
\begin{tabular}{c c c c c c c c c c}
\hline \hline 
& & \multicolumn{3}{c}{Experiment} & & \multicolumn{4}{c}{Calculated} \\
\cmidrule{3-5} \cmidrule{7-9} 
& & AF & UD15K & OD11K &&  & 6\% & 10\% & 14\% \\
\hline
\midrule

$a$ && 5.436 & 5.4 & 5.4 &&     & 5.3038 & 5.3028 & 5.292 \\
$b$ && 5.502 & 5.445 & 5.308 && & 5.4028 & 5.4080 & 5.406 \\
$c$ && 23.91 & 24.34 & 24.56 && & 24.5535 & 24.5568 & 24.5571 \\

\hline

\end{tabular}\label{table:latcon_ZZ}
\end{table}

\subsection{PBE calculations for the phonons}

\begin{table}
\caption{\label{tab:disp} Phonon frequencies from different functionals in meV. \textcolor{black}{Values obtained from the rev-vdW-DF2 functional are using the XY structure.}}

\begin{ruledtabular}
\scalebox{0.9}
{
\begin{tabular}{cccc}
$\vec{q}$ & PBE0 & PBE & rev-vdW-DF2 \\
\colrule
$\Gamma4$ & 4.233 & 4.897 & 0.421 \\
$\Gamma5$ & 4.272 & 5.129 & 0.423 \\
$\Gamma6$ & 5.489 & 5.698 & 0.945 \\
$\Gamma7$ & 6.297 & 6.776 & 0.977 \\
$\Gamma8$ & 7.667 & 7.330 & 0.979 \\
$\Gamma9$ & 8.572 & 8.073 & 1.111 \\
M5 & 9.005 & 9.009 & NA \\
M6 & 9.258 & 9.714 & NA \\
M7 & 9.438 & 9.877 & NA \\
M8 & 9.477 & 10.938 & NA \\
M9 & 9.957 & 11.517 & NA \\
$\Gamma10$ & 10.903 & 10.068 & 1.114 \\
M10 & 11.746 & 11.771 & NA \\
M11 & 12.024 & 12.236 & NA \\
M12 & 12.045 & 12.318 & NA \\
$\Gamma11$ & 12.152 & 10.463 & 1.647 \\
M13 & 12.396 & 12.521 & NA \\
M14 & 12.481 & 12.629 & NA \\
$\Gamma12$ & 12.712 & 10.836 & 1.660 \\
\end{tabular}
}
\end{ruledtabular}
\end{table}

\begin{figure*}
    \centering
    \includegraphics{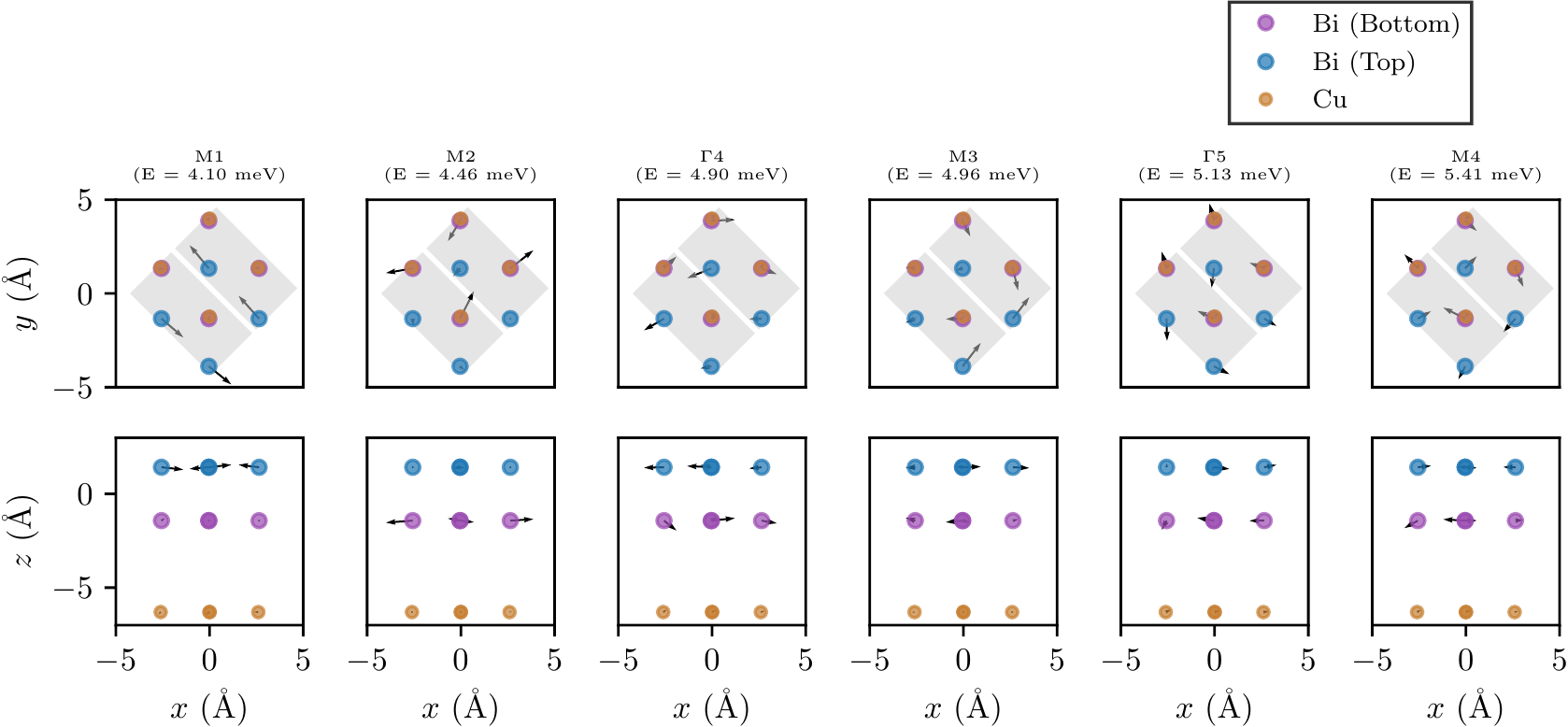}
    \caption{Phonon eigenvectors from DFT (PBE) calculations. Oxygen atoms are not included in the plots to clearly visualize the displacement of Bi atoms. Each column displays a phonon mode, with the top layer showing the top-down view and the bottom layer showing the side view.}
    \label{fig:eigvec_pbe}
\end{figure*}

\begin{figure}
    \includegraphics{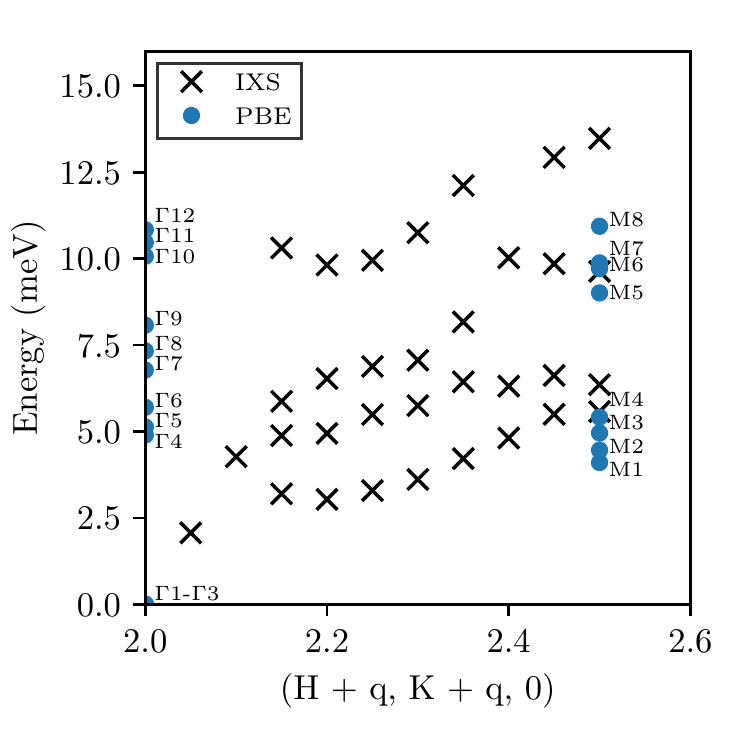}
    \caption{Theoretical and experimental phonon modes. The theoretical phonons were computed using DFT in the PBE approximation.} 
    \label{fig:disp_pbe}
\end{figure}

The PBE functional was also investigated. This approximation, however, did not converge to the starting antiferromagnetic ordering, which is a common issue in cuprate materials  \cite{horton2019high}. 
For this reason, most of the analysis and discussion in this work are done in the PBE0 functional. 
Nevertheless, the phonon frequencies in the PBE approximation are reported in Table~\ref{tab:disp} and are plotted in Fig.~\ref{fig:disp_pbe}.
The diagram of phonon eigenvectors is displayed in Fig.~\ref{fig:eigvec_pbe}. 
The M1 mode is associated with only the Bi atoms in the top plane moving horizontally, while the M2 mode shows only the Bi atoms in the bottom plane moving. 
The Bi planes are decoupled as shown in the bottom panels of Fig.~\ref{fig:eigvec_pbe}.

\bibliography{references}

\end{document}